\documentclass[reprint,amsmath,amssymb,aps,prl,superscriptaddress]{revtex4-1}

\usepackage{times,color,amsmath}
\usepackage{subfigure,epsfig,graphicx,epstopdf}
\usepackage[colorlinks]{hyperref}
\hypersetup{linkcolor = {blue},	citecolor = {blue},	urlcolor = {blue}}

\newcommand{\affa}{\affiliation{Center for Integrated Quantum Information Technologies (IQIT), School of Physics and Astronomy and State Key Laboratory of Advanced Optical Communication Systems and Networks, Shanghai Jiao Tong University, Shanghai 200240, China}}
\newcommand{\affb}{\affiliation{CAS Center for Excellence and Synergetic Innovation Center in Quantum Information and Quantum Physics, University of Science and Technology of China, Hefei, Anhui 230026, China}}
\newcommand{\affc}{\affiliation{Dipartimento di Fisica, Politecnico di Milano, Piazza L. da Vinci 32, I-20133 Milano, Italy}}
\newcommand{\affd}{\affiliation{IFISC (UIB-CSIC), Instituto de Fisica Interdisciplinar y Sistemas Complejos, E-07122 Palma de Mallorca, Spain}}
\newcommand{\affe} {\affiliation {TuringQ Co., Ltd., Shanghai 200240, China}}

\begin{document}

\title{ Experimentally Detecting Quantized Zak Phases without Chiral Symmetry in Photonic Lattices }

\author{Zhi-Qiang Jiao} \affa \affb   \author{Stefano Longhi} \affc \affd 
  \author{Xiao-Wei Wang}  \affa \affb	  \author{Jun Gao}  \affa \affb	
  
\author{Wen-Hao Zhou} \affa \affb          \author{Yao Wang} \affa \affb             \author{Yu-Xuan Fu} \affa
\author{Li Wang} \affa \affb	\author{Ruo-Jing Ren} \affa \affb \author{Lu-Feng Qiao} \affa \affb	\author{Xian-Min Jin} \email{xianmin.jin@sjtu.edu.cn}  \affa \affb \affe

\date{\today}

\begin{abstract}
 
Symmetries play a major role in identifying topological phases of matter and in establishing a direct connection between protected edge states and topological bulk invariants via the bulk-boundary correspondence. One-dimensional lattices are deemed to be protected by chiral symmetry, exhibiting quantized Zak phases and protected edge states, but not for all cases. Here, we experimentally realize an extended Su-Schrieffer-Heeger model with broken chiral symmetry by engineering one-dimensional zigzag photonic lattices, where the long-range hopping breaks chiral symmetry but ensures the existence of inversion symmetry. By the averaged mean displacement method, we detect topological invariants directly in the bulk through the continuous-time quantum walk of photons. Our results demonstrate that inversion symmetry protects the quantized Zak phase, but edge states can disappear in the topological nontrivial phase, thus breaking the conventional bulk-boundary correspondence. Our photonic lattice provides a useful platform to study the interplay among topological phases, symmetries, and the bulk-boundary correspondence.
\end{abstract}
\maketitle

Topological phases of matter are fascinating states  that escape from the standard description of Ginzburg-Landau theory, exhibiting protected edge states and quantized topological properties in the bulk\cite{integer_Hall, SSH, short_course, protect,symmetry}. A paradigm of topological phases in condensed-matter physics is the integer quantum Hall state in a two-dimensional electron gas \cite{integer_Hall}. Topological order is of major relevance  in a variety of nonelectronic systems as well, including mechanical platform, cold atoms, acoustics, trapped ions and photonics \cite{mechanics, sound, cold_gas, acoustic, photonics, Wang, correlation, photonics_RMP, AB, AB_experiment}. A central result in the theory of topological matter is that bulk topological invariants are connected to the number of protected edge states (bulk-boundary correspondence) \cite{short_course,symmetry}. Hence topological phases can be probed either in the bulk or at the edges, as demonstrated in several recent experiments for Hermitian and even some non-Hermitian systems \cite{Kitagawa, quasicrystal, Floquet, Hafezi, driven_lattice, anomalous, Pan2018, 4D_Hall,invariants, graphene, transition, Cardano, disordered, chiral, midgap, robust_transition, Xue, Chong, Zero_mode}. 
However, the common belief  that a nontrivial topological phase implies the presence of protected edge states is not a general result, and even in Hermitian systems it is known that there could exist nontrivial topological phases that do not exhibit edge states at all \cite{short_course,new1,semimetal,Platero}.
In such systems, edge dynamics alone cannot thus provide a safe diagnostic of a topological phase. A notable example is provided by one-dimensional (1D) and 2D lattices with broken chiral symmetry but with preserved inversion symmetry \cite{short_course,semimetal,Platero,Song,proposal,cazzo1,cazzo2}.
 In the Altland-Zirnbauer (AZ) classification, nontrivial topological phases are identified by three main symmetries: time-reversal, chiral and particle-hole symmetries \cite{chiral_importance,symmetry}.
In this scheme,  a topologically nontrivial 1D system should possess chiral symmetry, and the bulk topological invariant is the winding number corresponding to a quantized Zak phase \cite{Zak}.  A paradigmatic example is provided by the famous Su-Schrieffer-Heeger (SSH) model of polyacetylene \cite{SSH, short_course}, where the bulk-boundary correspondence holds \cite{ short_course,new2}. Experimental demonstrations of quantized Zak phases in the SSH model with chiral symmetry have been reported in recent works \cite{measure_Zak,chiral} using different platforms and methods to extract the Zak phase from bulk dynamics. Long-range hopping  in the SSH model that breaks chiral symmetry makes the system topologically trivial.  However, when inversion symmetry is preserved, the Zak phase remains quantized and one can introduce a $Z_2$ topological number which identifies two distinct topological phases, which are not caught by the AZ classification \cite{semimetal,Platero,Song}. Interestingly, unlike the chiral SSH model, in the extended SSH model with inversion symmetry only, the quantized Zak phase can no longer predict the number of edge states since bulk-boundary correspondence cannot be established under inversion symmetry solely \cite{short_course}.\\

 In this Letter, we report on the experimental demonstration of quantized Zak phases in the extended SSH model with broken chiral symmetry and demonstrate that the bulk topological invariant cannot predict the number of edge states.  
\begin{figure}[!t]
	\centering
	\includegraphics[width=0.98\linewidth]{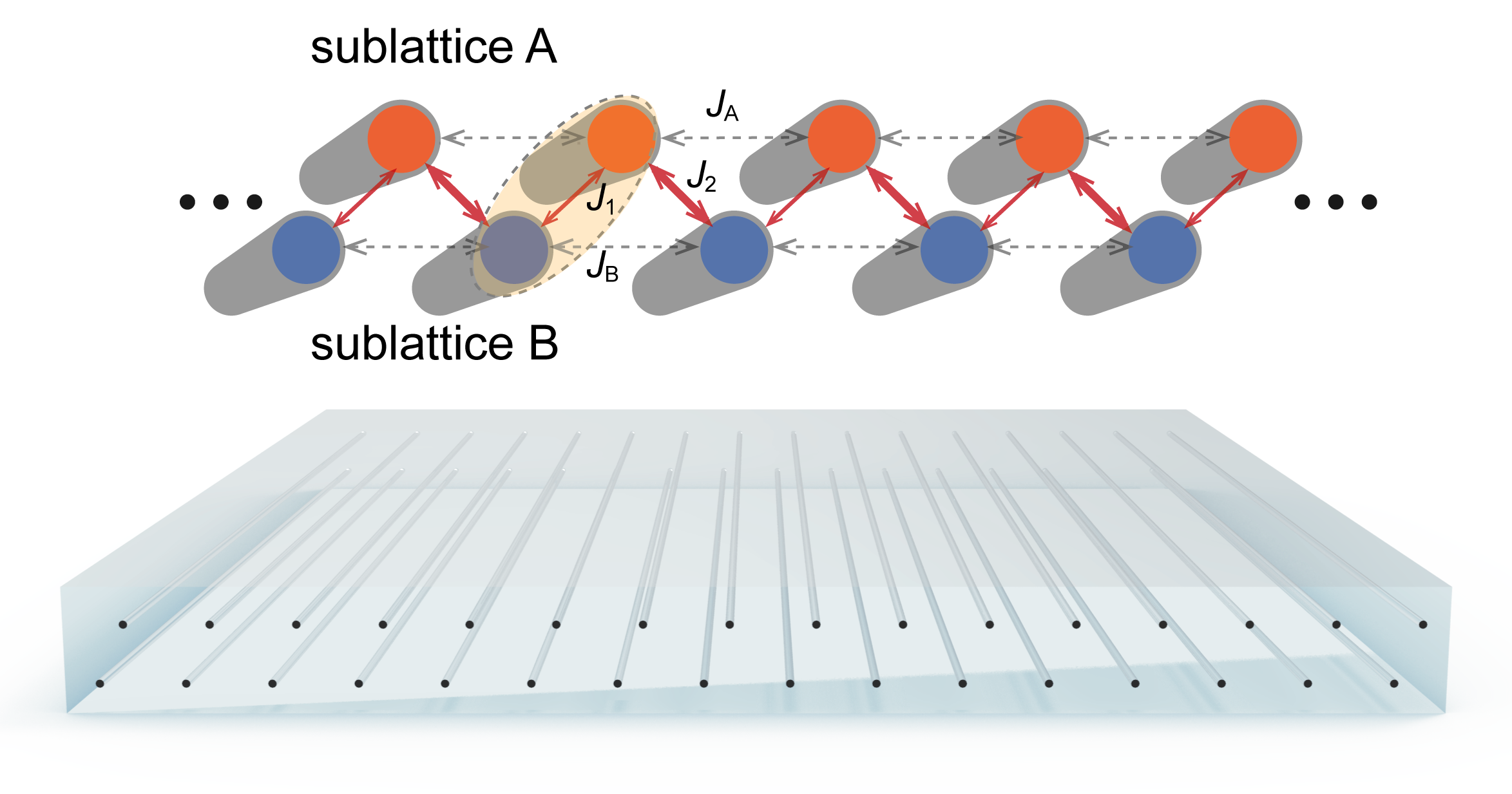}
	\caption{\textbf {Extended SSH lattice.}  Schematic of a bipartite lattice in a zigzag geometry, comprising two sublattices A and B with staggered coupling constants $J_1$ and $J_2$ and with next nearest neighbor couplings (NNNCs) $J_{A}$ and $J_{B}=J_A$. The NNNCs break the chiral symmetry of the lattice while preserving the inversion symmetry.}			
	\label{fig1}
\end{figure}
The two-band insulating system is built on a photonic chip under a continuous-time quantum walk framework.  Waveguide lattices are a natural platform  to study topological properties, both in the bulk and at the edges, in virtue of photon dynamics  \cite{quasicrystal,Floquet,SupplementalMaterials,Hamiltonian,review_QW,2DQW}. The quantized Zak phase can be directly detected from bulk dynamics by means of the beam displacement method \cite{chiral, proposal}, while edge state dynamics can be visualized by boundary excitation of the lattice \cite{Floquet,proposal}. We constructed an extended SSH lattice using a zigzag design to introduce weak next nearest neighbor couplings (NNNCs) in order to break the chiral symmetry \cite{Platero}, as shown in Fig.\ref{fig1}. By tuning the intercell and intracell hopping amplitudes, we observe  quantization of the Zak phase. The nontrivial topological edge states are initially protected by inversion symmetry. However, as the NNNCs increases to a level higher than the half of the intercell hopping strength (while the intracell hopping remains unchanged), edge states delocalize and their energies flow inside the photonic lattice bands, despite the bulk topological invariant is unchanged and non-vanishing \cite{Platero,Song,proposal}.

\begin{figure*}[!t]
	\centering
	\includegraphics[width=1.0\linewidth]{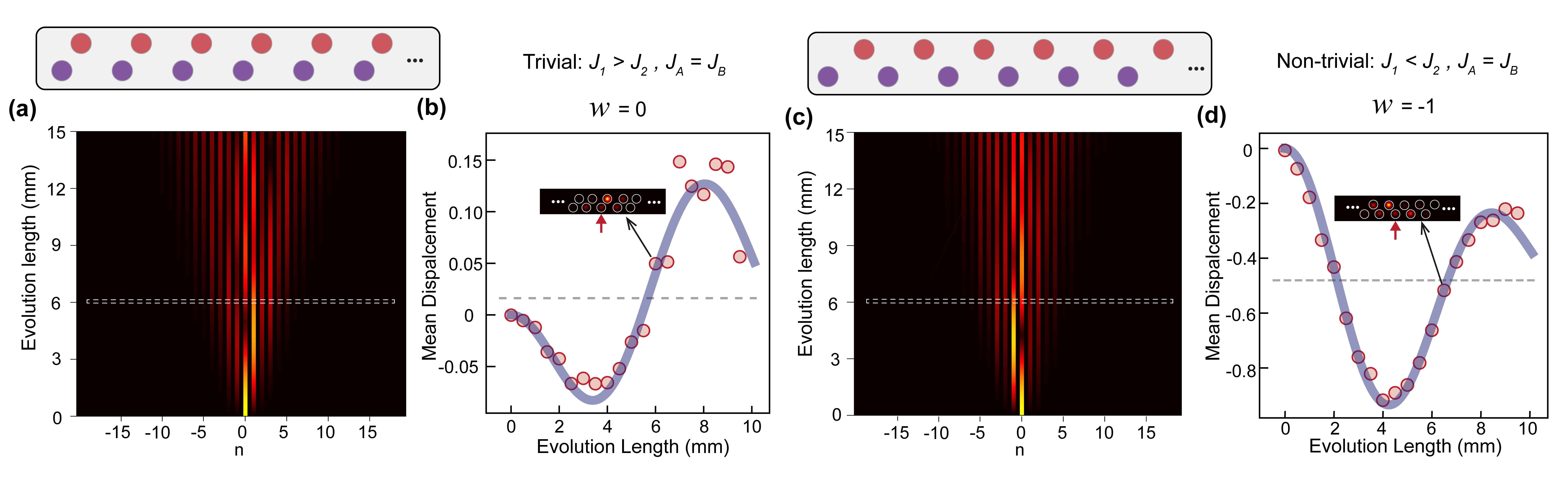}
	\caption{
		\textbf{Averaged mean displacement and bulk topological number.} \textbf{(a)} (\textbf{(c)}) is the simulated light intensity dynamics in the bulk excitation of trivial (nontrivial) topological lattices. The light intensities of each waveguide are displayed versus the propagation distance $z$ on a pseudo color map. The topological trivial phase $(\mathcal{W}=0$) corresponds to $J_{1} > J_{2}$, while $J_{1} < J_{2}$ corresponds to the topological nontrivial phase ($\mathcal{W}=-1$). 
		\textbf{(b)} and \textbf{(d)} are the averaged mean displacements in trivial and nontrivial cases. The solid  blue curves show the theoretical fitting curves of the spatial beam displacement $P(z)$ versus $z$. The experimental results are labeled by red dots. Every dot corresponds to the mean beam displacement obtained from the 40-site recored images (examples at $z=6.5$ mm are shown in insets of figures). The averaged mean displacement of trivial (nontrivial) lattice is 0.005 (-0.455), highlighted by the gray horizontal dashed line. 
		 }
	\label{fig2}
\end{figure*}

 The SSH model is composed by a one-dimensional $N$-site lattice of even number with staggered hopping coefficients $J_1$ and $J_2$. Generally, it can be divided into two sublattices, odd sites $A = \{1, 3, ..., N-1\}$ and even sites $B = \{2, 4, ..., N\}$. 
As shown Fig.\ref{fig1}, we realize an extended SSH model with NNNCs $J_A, J_B$ using a binary zigzag waveguide arrangement. The long-range hopping in sublattices A and B breaks the chiral, but not the inversion symmetry provided that $J_A=J_B$ \cite{semimetal,Platero}. The tight-binding Hamiltonian of the photonic lattice reads 
\cite{photonics_RMP},
\begin{equation}
H= \sum_{n}(J_{1}a^{\dagger}_{n}b_{n}+J_{2}b^{\dagger}_{n}a_{n+1}+J_{A}a^{\dagger}_{n}a_{n+1}+J_{B}b^{\dagger}_{n}b_{n+1}) + H.c.
\label{Eq1}
\end{equation}
where $a^{\dagger}_{n}$ ($a_n$) and $b^{\dagger}_{n}$ $(b_{n})$ are the creation (annihilation) operators of photon fields in the various guides of the two sublattices.
The temporal evolution of the insulating system can be mapped into the longitudinal photon propagation in the waveguide lattice.
The bulk Hamiltonian in moment space can be written in terms of Pauli matrices as \cite{Platero,proposal}
\begin{equation}
\mathcal{H}(k) = \left( 
\begin{matrix}
	c_{A}(k) & c_{1}(k) \\
	c^{\ast}_{1}(k) & c_{B}(k)
\end{matrix}
\right)    \\ = \frac{c_{A}(k)+c_{B}(k)}{2}\sigma_{0} + \mathbf{h}(k) \cdot \pmb{\sigma}
\label{Eq2}
\end{equation}
where: $c_{A,B}(k) = 2J_{A,B} \cos(k)$ describe the hopping among the sites of sublattice A and B, respectively; $c_{1}(k) = J_{2} +J_{1} \exp(-ik) \equiv H(k) \exp[ i \phi(k)]$ describes the hopping among sites in different sublattices; $k$ is the quasimomentum ranging from $-\pi$ to $\pi$;  $\sigma_{0}$ is the identity matrix; $\pmb{\sigma}$ is the vector form of Pauli matrices ($\sigma_{x}$, $\sigma_{y}$, $\sigma_{z}$); and $\mathbf{h}(k)$ is defined by the relation 
\begin{equation}
\mathbf{h}(k) = (H \cos\phi,-H \sin\phi,\frac{c_{A}-c_{B}}{2}).
\label{Eq3}
\end{equation}   

The system displays chiral ($\Gamma$) and inversion ($\mathcal{R}$) symmetry whenever $\Gamma \mathcal{H}(k) \Gamma^{-1} = - \mathcal{H}(k)$ and $\mathcal{R} \mathcal{H}(k) \mathcal{R}^{-1} = \mathcal{H}(-k)$ \cite{symmetry}. For a two-band system, the two symmetries are defined by the unitary operators  $\Gamma=\sigma_{z}$  and $\mathcal{R}=\sigma_{x}$. Thus a lattice with chiral symmetry requires $c_{A}(k)=c_{B}(k) = 0$, i.e. $J_A=J_B=0$, while inversion symmetry only requires $J_A=J_B$. 
As a result, an equal NNNCs strength inside sublattices makes the chiral symmetry broken, while inversion symmetry is preserved.
 The Zak phase, that is, the Berry phase accumulated by an eigenstate adiabatically transporting through the whole Brillouin zone, is a bulk property of the insulating system that is quantized in the presence of chiral symmetry. When the chiral symmetry is broken but the system still possesses inversion symmetry, i.e. for $J_A=J_B$,  the NNNCs terms modify the energy spectrum of the bulk Hamiltonian (2) but not the corresponding eigenfunctions, so that quantization of the Zak phase is preserved by inversion symmetry. In this case the Zak phase in the two bands takes the same value, given by 
\begin{equation}
\gamma = \frac{1}{2}\int_{-\pi}^{\pi} dk\frac{\partial \phi}{\partial k} = \pi \mathcal{W}
\label{Eq.4}
\end{equation}
where $\mathcal{W} = \frac{1}{2 \pi} \int_{-\pi}^{\pi} dk(\mathbf{n}\times\frac{\partial \mathbf{n}}{\partial k})$ is the winding number associated to the motion of the vector $\mathbf{n}(k)=(\cos \phi, -\sin \phi,0)$. 

For a chiral system, $\gamma=0$ corresponds to the topological trivial phase ($J_1>J_2$) without edge states in a chain with open boundaries, while $| \gamma |=\pi$ corresponds to the nontrivial topological phase ($J_1<J_2$) with two zero-energy edge states (which do not hybridize in the thermodynamic limit). To change $| \gamma |$  from $\pi$ to $0$, i.e. to remove edge states, one needs to close and reopen the gap at $J_1=J_2$, where the two bands touch at $k= \pm \pi$. For a system with broken chiral symmetry but with inversion symmetry, the eigenfunctions (and thus the Zak phase $\gamma$) are independent of the NNNCs,  however the energy bands depend on the NNNC and gap closing can occur for a strong NNNC without bound touching (owing to the indirect nature of the gap \cite{Platero,proposal}). As a consequence, edge states in the topological nontrivial phase $|\gamma|=\pi$ can delocalize and flow inside the bands as the NNNCs is increased \cite{semimetal,Platero,proposal}, which is impossible in an insulating system with chiral symmetry owing to the bulk-boundary correspondence. 

To measure the Zak phase in our photonic platform, we use the mean displacement method \cite{chiral,proposal}, looking at the asymptotic mean spatial displacement of a time-evolved wave packet corresponding to a single-site excitation in the bulk of the lattice. We define the mean wave packet displacement at propagation distance $z$ (corresponding to evolution in time $t=z/c$) as $P(z) = \sum_{n} n(|A_{n}(z)|^{2} + |B_{n}(z)|^{2})$, where $A_{n}(z)$ ($B_{n}(z)$) are the field amplitudes in sublattice A (B) in the $n$-th cell. The averaged mean wave packet displacement in the interval $(0,Z)$ is defined by 
\begin{equation}
\overline{P}(Z) = \frac{1}{Z} \int_{0}^{Z} P(z) dz.
\label{eq4}
\end{equation}
For large $Z$,  $\overline{P}(Z)$ approaches a constant asymptotic value \cite{chiral,proposal}. Remarkably,
when the two-band system displays inversion symmetry, one has $\mathcal{W} = 2\overline{P}(Z)$ for large $Z$ \cite{proposal}. Hence a measurement of the mean beam displacement provides a  measure of the winding number $\mathcal{W}$, and thus of the Zak phase $\gamma$. \par

The binary waveguide lattice in the zigzag geometry shown in Fig.\ref{fig1} is manufactured on a photonic chip by the femtosecond laser direct-writing technique  \cite{fs-laser_writing, 2nd_coupling, Bloch_oscilations}. 
We prepare four photonic chips to realize four kinds of lattices in different topological phases and tuned NNNCs strengths. Each array comprises 20 unit cells with open boundaries.  The hopping coefficients are determined by the evanescent mode coupling between adjacent waveguides, which can be tailored by controlling waveguide spacing \cite{SupplementalMaterials}. Once the hopping parameters are fixed, the geometry structure of the lattice can be easily generated. The first chip realizes a topological trivial insulator with the hopping coefficients $J_{1}= 0.30$, $ J_{2}=0.13$, and $J_{A} =J_{B}=0.05$ (in units of 1/mm), corresponding to intercell distance 13.2 $\mu$m, intracell distance 11.2 $\mu$m, and NNNCs distance 16.5 $\mu$m.
 The second chip realizes a topological nontrivial lattice, with intercell distance 10.7 $\mu$m,  intracell distance 14.4 $\mu$m, and NNNCs distance 16.5 $\mu$m, corresponding to $J_{1}=0.09$, $J_{2}=0.36$ and $J_{A} =J_{B}=0.05$ (in units of 1/mm). In both cases, the weak NNNCs ensures that the system is an insulator with a direct open gap. To calculate the averaged mean displacement, we fabricate the photonic lattices with a series of evolution lengths increasing gradually from 0 mm to 9.5 mm with an interval of 0.5 mm. The arrays are probed by exciting the central waveguide with a coherent laser at 780 nm wavelength, and the intensity distributions at various output planes are recorded on a CMOS camera (see \cite{SupplementalMaterials} for details). The results are depicted in Fig.2. \begin{figure*}[!t]
	\centering
	\includegraphics[width=1.0\linewidth]{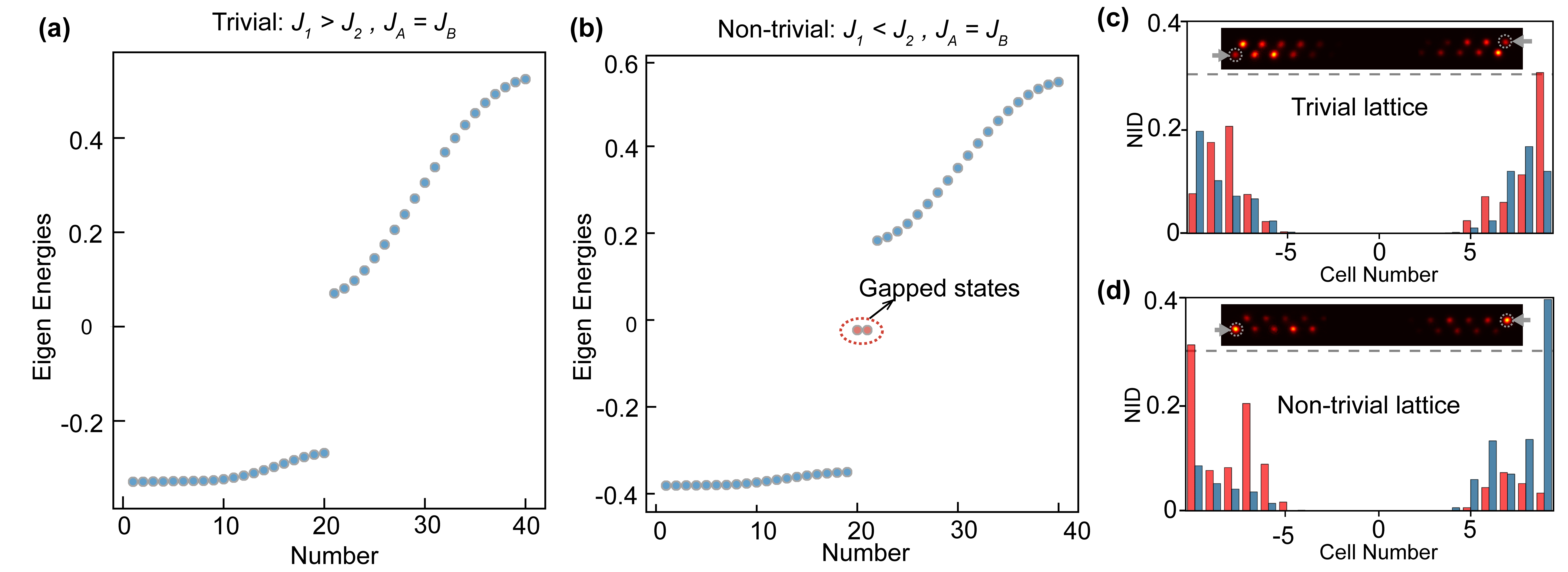}
	\caption{
	\textbf{The energy spectrum and  edge dynamics.} \textbf{(a)} and \textbf{(b)} shows the energy spectrum of the trivial and nontrivial lattices, respectively. In \textbf{(a)} there is an energy gap between the two bands without any gapped state, while in \textbf{(b)} there are two gapped (edge) states, partly protected by inversion symmetry. \textbf{(c)} and \textbf{(d)} are the normalized intensity distributions (NID) of the 40-site lattices in the trivial and nontrivial phases, respectively, as measured at the output plane of the chips. The photonic lattices are excited in the two edges labeled by the dashed gray circles.}		
	\label{fig3}
\end{figure*}
Figures \ref{fig2} (a) and (c) show the numerically computed light intensity distribution versus propagation distance $z$ in the topologically trivial (a) and nontrivial (c) phases for a evolution length of 15 mm.
 The mean displacement $P{(z)}$ versus $z$ is obtained from the recorded intensity distributions at various propagation distances [insets in Figs.2(b) and (d)]. The experimental results are shown in Figs. \ref{fig2} (b) and (d). In these plots, the red dots refer to the experimental measurements of $P(z)$, while the blue solid curves correspond to the simulated results obtained from the intensity maps of Figs.\ref{fig2} (a) and (c).  The averaged mean displacements $\overline{P}(Z)$ at $Z=9.5$ mm, obtained from the fitting theoretical curves of $P(z)$ and using Eq.(\ref{eq4}), are depicted by the horizontal gray dashed lines in Figs.\ref{fig2}(b) and (d). In the topological trivial lattice [panel (b)], one obtains $\overline{P}(Z)=0.001 \pm 0.05$, in good agreement with the theoretical winding number $\mathcal{W}=0$. In the topological nontrivial lattice [panel (d)], a winding number $\mathcal{W}  \simeq -1$ can be estimated from  the measured mean displacement value $\overline{P}(Z) = -0.455 \pm 0.08$. Uncertainties in $\overline{P}(Z)$, indicated by the $\pm$ error bars, are estimated by assuming possible error in the coupling constants of the lattices, arising from fabrication imperfections of the geometric structure, that can slightly deviate from the target values and thus change the fitting curves $P(z)$ and the corresponding mean value $\bar{P}(Z)$.

\begin{figure}
	\centering
	\includegraphics[width=1.0\linewidth]{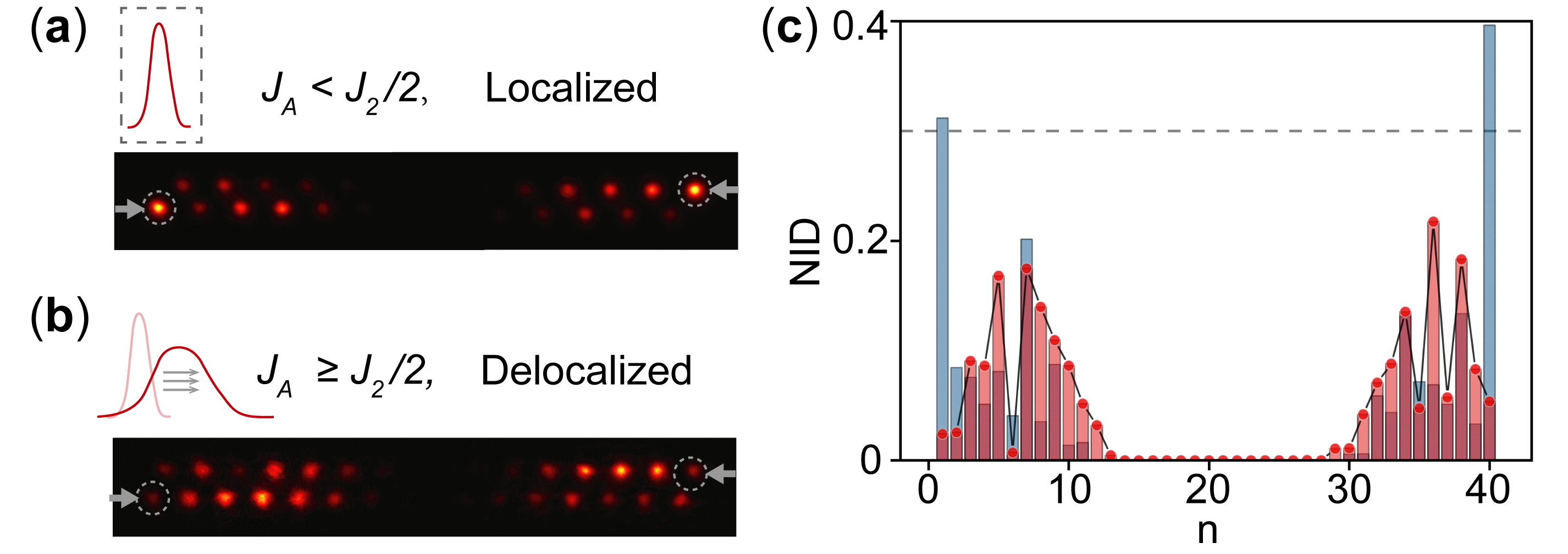}
	\caption{ \textbf{Delocalization of edge states in the topological phase.}
	 \textbf{(a,b)} Intensity light patterns at the output facets of nontrivial topological lattices, initially excited at the two edges, imaged by a CMOS camera. In (a)    $J_A<J_2/2$ (open gap), while in (b) $J_A>J_2/2$ (closed gap).  
	 \textbf{(c)} Normalized light intensity distribution (NID) versus lattice site $n$ \textbf{(a)} (blue) and \textbf{(b)} (red) after $Z=18$ mm propagation distance. For the strong NNNCs case the energy gap closes, causing the light to delocalize in the bulk of the lattice (shown in red columns)}.
	\label{fig4}
\end{figure}

To study edge dynamics, we excited the two lattices at the left/right edge waveguides and recorded the intensity distributions at the output facets of the chips, as shown in Fig.\ref{fig3} (c) and (d). For a topological trivial lattice, there are no edge states and the energy spectrum, shown in Fig.\ref{fig3}(a), is formed by two bands without gapped states; this results in a clear delocalization of light far from the initially excited boundary waveguides, as shown in Fig.\ref{fig3}(c). Note that the energy spectrum is not symmetric around $E=0$ because of the broken chiral symmetry introduced by the NNNCs. In the topological nontrivial lattice, two gapped states, corresponding to two localized edge states at the left and right boundaries of the lattice, clearly emerge in the energy spectrum [Fig.\ref{fig3}(b)], resulting in a relative localization of excitation at the edge waveguides observed in the intensity map of Fig.\ref{fig3}(d).
As the NNNCs $J_A=J_B$ is increased above $J_2/2$, the gap between the two bands becomes indirect and can close without band touching \cite{Platero,proposal}; correspondingly, the two gapped states in the nontrivial topological phase delocalize and flow inside the bands. In this strong NNNCs regime, we do not observe any edge state even though the bulk topological invariant $\mathcal{W}$ is nonvanishing. We experimentally demonstrated this effect by manufacturing two   chips of the 18-mm-long evolution length both in the topological nontrivial phase, with $J_{A}= J_{B}=0.2$, $J_{1} = 0.15 $, $J_{2}=0.3$ and $J_{A}= J_{B}=0.05$, $J_{1} =0.09$, $J_{2}=0.36$. In the former case ($J_A>J_2/2$), the gap is closed and thus there are no edge states, while in the latter case  ($J_A<J_2/2$) the gap is open and there are two edge states.
The edge dynamics in the two lattices is shown in Fig.\ref{fig4}.
The experimental results clearly show the disappearance of the edge states in the strong NNNCs regime $J_A>J_2/2$. Since the winding number $\mathcal{W}$ does not depend on $J_A$, the disappearance of the edge states does not change the bulk topology of the lattice, which remains nontrivial ($\mathcal{W}=-1$). Indeed, we measured the winding number for the lattice with  $J_A>J_2/2$ and clearly found a non-vanishing winding (see Fig.S4 in the Supplemental Materials \cite{SupplementalMaterials}). This result demonstrates that edge dynamics alone is insufficient to characterize the topological properties of the bulk Hamiltonian in the extended SSH model.

In conclusion, we have experimentally demonstrated on a photonic waveguide platform that an 1D two-band insulating system with inversion symmetry, but with broken chiral symmetry, can exhibit topological nontrivial phases, characterized by a quantized Zak phase. Such topological nontrivial phases go beyond the standard AZ classification and show that a nontrivial topological bulk phase does not necessarily correspond to the existence of edge states. Our experimental platform of synthetic topological matter enabled us to perform both bulk measurements, to detect the band topological invariant (Zak phase) and edge measurements, and to detect edge states. It could be feasible to similarly explore other topological models---for example parity-time ($\mathcal{PT}$) symmetry or non-Hermitian systems without chiral symmetry \cite{proposal} and to measure  fractional winding numbers and test bulk-boundary correspondence in non-Hermitian models \cite{Lee}.

\begin{acknowledgments}
	The authors thank Jian-Wei Pan for helpful discussions. This research is supported by National Key R\&D Program of China (2019YFA0308700, 2019YFA0706302 and 2017YFA0303700), National Natural Science Foundation of China (NSFC) (11904229, 11761141014, 61734005, 11690033), Science and Technology Commission of Shanghai Municipality (STCSM) (20JC1416300, 2019SHZDZX01), Shanghai Municipal Education Commission (SMEC)(2017-01-07-00-02-E00049). X.-M.J. acknowledges additional support from a Shanghai talent program. X.-M.J. acknowledges support from a Shanghai talent program and support from Zhiyuan Innovative Research Center of Shanghai Jiao Tong University.
\end{acknowledgments}


\begin{thebibliography}{99}
\bibitem{integer_Hall}K. v. Klitzing, G. Dorda, M. Pepper, New method for high-accuracy determination of the fine-structure constant based on quantized Hall resistance. \textit{Phys. Rev. Lett.} \textbf{45}, 494-497 (1980).	
\bibitem{SSH}W. P. Su, J. R. Schieffer, and A. J. Heeger. Solitons in Polyacetylene.\textit{Phys. Rev. Lett.} \textbf{42}, 1698 (1979).
\bibitem{protect} B. I. Halperin, Quantized Hall conductance, current-carrying edge states, and the existence of extended states in a two-dimensional disordered potential. \textit{Phys. Rev. B} \textbf{25}, 2185 (1982).
\bibitem{short_course}J.K. Asb\'{o}th, L. Oroszl\'{a}ny,and A. P\'{a}lyi, A short course on topological insulators, \textit{Lect. Notes Phys.} \textbf{919}, 1 (2016).
\bibitem{symmetry} C.-K. Chiu, J.C.Y. Teo, A.P. Schnyder, and S. Ryu, Classification of topological quantum matter with symmetries,
\textit{Rev. Mod. Phys.} {\bf 88}, 035005 (2016).
\bibitem{mechanics}S. D. Huber, Topological mechanics, \textit{Nat. Phys.} \textbf{12}, 621 (2016).
\bibitem{sound} V. Peano, C. Brendel, M. Schmidt, F. Marquardt, Topological phases of sound and light. \textit{Phys. Rev. X} \textbf{5}, 031011 (2015).
\bibitem{cold_gas} N. Goldman, J. C. Budich, P. Zoller, Topological quantum matter with ultracold gases in optical lattices. \textit{Nat. Phys.} \textbf{12}, 639 (2016).
\bibitem{acoustic}M. Xiao, G.-C. Ma, Z.-Y. Yang, P. Sheng, Z.-Q. Zhang, C.-T. Chan, Geometric phase and band inversion in periodic acoustic systems. \textit{Nat. Phys.} \textbf{11}, 240 (2015).
\bibitem{photonics} L. Lu, J. D. Joannopoulos, M. Solja\v{a}ic, Topological photonics. \textit{Nat. Photon.} \textbf{8} 821-829 (2014).
\bibitem{photonics_RMP}  T. Ozawa, H. M. Price, A. Amo, N. Goldman, M. Hafezi, L. Lu, M. C. Rechtsman, D. Schuster, J. Simon, O. Zilberberg, I. Carusotto, Topological photonics. \textit{Rev. Mod. Phys.} \textbf{91}, 015006 (2019).
\bibitem{AB} S. Longhi, Aharonov–Bohm photonic cages in waveguide and coupled resonator lattices by synthetic magnetic fields. \textit{Opt. Lett.} \textbf{39}, 5892-5895 (2014)
\bibitem{AB_experiment} S. Mukherjee, M. Di Liberto, P. \"{O}hberg, R. R. Thomson, N. Goldman. Experimental observation of Aharonov-Bohm cages in photonic lattices. \textit{Phys. Rev. Lett.} \textbf{121}, 075502 (2018).
\bibitem{Wang} Y. Wang, Y.-H. Lu, F. Mei, J. Gao, Z.-M. Li, H. Tang, S.-L. Zhu, S.-T. Jia,  X.-M. Jin, Direct observation of topology from single-photon dynamics. \textit{Phys. Rev. Lett.} \textbf{122}, 193903(2019).
\bibitem{correlation}Y. Wang, X.-L. Pang, Y.-H. Lu, J. Gao, Y.-J. Chang, L.-F. Qiao, Z.-Q. Jiao, H. Tang, and X.-M. Jin, Topological protection of two-photon quantum correlation on a photonic chip. Optica \textbf{6}, 955 (2019).
\bibitem{Kitagawa}  T. Kitagawa, M. A. Broome, A. Fedrizzi, M. S. Rudner, E. Berg, I. Kassal, A. Aspuru-Guzik, E. Demler,  A. G. White, Observation of topologically protected bound states in photonic quantum walks. \textit{Nat. Commun.} \textbf{3}, 882 (2012).
\bibitem{quasicrystal} Y. E. Kraus, Y. Lahini, Z. Ringel, M. Verbin, and O. Zilberberg, Topological states and adiabatic pumping in quasicrystals, \textit{Phys. Rev. Lett.} \textbf{109}, 106402 (2012).
\bibitem{Floquet} M. C. Rechtsman, J. M. Zeuner, Y. Plotnik, Y. Lumer, D. Podolsky, F. Dreisow, S. Nolte, M. Segev,  A. Szameit, Photonic Floquet topological insulators. \textit{Nature} \textbf{496}, 196 (2013).
\bibitem{Hafezi}M. Hafezi, S. Mittal, J. Fan, A. Migdall, J. M. Taylor, Imaging topological edge states in silicon photonics. \textit{Nat. Photon.} \textbf{7}, 1001–1005 (2013).
\bibitem{driven_lattice} S. Mukherjee, A. Spracklen, M. Valiente, E. Andersson, P. \"{O}hberg, N. Goldman, R. R. Thomson, Experimental observation of anomalous topological edge modes in a slowly driven photonic lattice. \textit{Nat. Commun.} \textbf{8}, 13918 (2017).
\bibitem{anomalous}  L. J. Maczewsky, J. M. Zeuner, S. Nolte, A. Szameit, Observation of photonic anomalous Floquet topological insulators. \textit{Nat. Commun.} \textbf{8}, 13756 (2017).
\bibitem{Pan2018}C. Chen, X. Ding, J. Qin, Y. He, Y.-H. Luo, M.-C. Chen, C. Liu, X.-L. Wang, W.-J. Zhang, H. Li, L.-X. You, Z. Wang, D.-W. Wang, B.C. Sanders, C.-Y. Lu, J.-W. Pan, Observation of topologically protected edge states in a photonic two-dimensional quantum walk. \textit{Phys. Rev. Lett.} \textbf{121}, 100502 (2018).
\bibitem{4D_Hall}  O. Zilberberg, S. Huang, J. Guglielmon, M. Wang, K. P. Chen, Y. E. Kraus, M. C. Rechtsman, Photonic topological boundary pumping as a probe of 4D quantum Hall physics. \textit{Nature} \textbf{553}, 59 (2018).
\bibitem{invariants} M. Hafezi, Measuring topological invariants in photonic systems, \textit{Phys. Rev. Lett.} \textbf{112}, 210405 (2014).
\bibitem{graphene} Y. Plotnik, M. C. Rechtsman, D.-H. Song, M. Heinrich, J. M. Zeuner, S. Nolte, Y. Lumer, N. Malkova, J.-J. Xu, A.r Szameit, Z-G. Chen, M. Segev, Observation of unconventional edge states in photonic graphene, \textit{Nat. Mat.} \textbf{13}, 57 (2014)
\bibitem{transition} J. M. Zeuner, M. C. Rechtsman, Y. Plotnik, Y. Lumer, S. Nolte, M. S. Rudner, M. Segev, A. Szameit, Observation of a topological transition in the bulk of a non- Hermitian system, \textit{Phys. Rev. Lett.} \textbf{115}, 040402 (2015).
\bibitem{Cardano}  F. Cardano, M. Maffei, F. Massa, B. Piccirillo, C. de Lisio, G. De Filippis, V. Cataudella, E. Santamato, L. Marrucci, Statistical moments of quantum-walk dynamics reveal topological quantum transitions, \textit{Nat. Commun.} \textbf{7}, 11439 (2016).
\bibitem{disordered}S. Barkhofen, T. Nitsche, F. Elster, L. Lorz, A. G\'{a}bris, I. Jex, C. Silberhorn, Measuring topological invariants in disordered discrete time quantum walks, \textit{Phys. Rev. A} \textbf{96}, 033846 (2017).
\bibitem{chiral} F. Cardano, A. D’Errico, A. Dauphin, M. Maffei, B. Piccirillo, C. de Lisio, G. De Filippis, V. Cataudella, E. Santamato, L. Marrucci, M. Lewenstein, P. Massignan, Detection of Zak phases and topological invariants in a chiral quantum walk of twisted photons, \textit{Nat. Commun.} \textbf{8}, 15516 (2017).
\bibitem{midgap} H. Schomerus, Topologically protected midgap states in complex photonic lattices, \textit{Opt. Lett.} \textbf{38}, 1912 (2013).
\bibitem{robust_transition} H. Zhao, S. Longhi, and L. Feng, Robust light state by quantum phase transition in non-Hermitian optical materials, \textit{Sci. Rep.} \textbf{5}, 17022 (2015).
\bibitem{Xue}  L. Xiao, X. Zhan, Z.-H. Bian, K.-K. Wang, X. Zhang, X.-P. Wang, J. Li, K. Mochizuki, D. Kim, N. Kawakami, W. Yi, H. Obuse, B. C. Sanders, P. Xue, Observation of topological edge states in parity-time-symmetric quantum walks, \textit{Nat. Phys.} \textbf{13}, 1117 (2017).
\bibitem{Chong} D. Leykam, K. Y. Bliokh, C. Huang, Y.D. Chong, F. Nori, Edge modes, degeneracies, and topological numbers in non-hermitian systems, \textit{Phys. Rev. Lett.} \textbf{118}, 040401 (2017)
\bibitem{Zero_mode} M. Pan, H. Zhao, P. Miao, S. Longhi, L. Feng, Photonic zero mode in a non-Hermitian photonic lattice, \textit{Nat. Commun.} \textbf{9}, 1308 (2018).
\bibitem{new1}
T.L. Hughes, E. Prodan, and B. A. Bernevig, Inversion-symmetric topological insulators,
\textit{Phys. Rev. B} {\bf 83}, 245132 (2011).
\bibitem{semimetal} G. van Miert, C. Ortix and C. Morais Smith, Topological origin of edge states in two-dimensional inversion-symmetric insulators and semimetals, \textit{2D Mater.} \textbf{4}, 015023 (2017).

\bibitem{Platero}
B. Perez-Gonzalez, M. Bello, A. Gomez-Leon, and G. Platero, Interplay between long-range hopping and disorder in topological systems,
\textit{Phys. Rev. B}  {\bf 99}, 035146 (2019).

\bibitem{Song} B. Song, L. Zhang, C.-D. He, T. F. J. Poon, E. Hajiyev, S.-C. Zhang, X.-J. Liu, G.-B. Jo, Observation of symmetry-protected topological band with ultracold fermions, \textit{Sci. Adv.} \textbf{4}, eaao4748 (2018).
\bibitem{proposal} S. Longhi, Probing one-dimensional topological phases in waveguide lattices with broken chiral symmetry, \textit{Opt. Lett.} \textbf{43}, 4639 (2018).

\bibitem{cazzo1}
M. Di Liberto, D. Malpetti, G. I. Japaridze, and C. Morais Smith, Ultracold fermions in a one-dimensional bipartite optical lattice: Metal-insulator transitions driven by shaking, \textit{Phys. Rev. A} {\bf 90}, 023634 (2014).

\bibitem{cazzo2}
S.R. Pocock, P.A. Huidobro, and V. Giannini,
Bulk-edge correspondence and long-range hopping in the topological plasmonic chain, \textit{Nanophoton.} {\bf 8},  1337 (2019).

\bibitem{chiral_importance}S. Ryu, A.P. Schnyder, A. Furusaki, A.W.W. Ludwig, Topological insulators and superconductors: ten-fold way and dimensional hierarchy, \textit{New J. Phys.} \textbf{12}, 065010 (2010).
\bibitem{Zak} J. Zak, Berry phase for energy bands in solids. \textit{Phys. Rev. Lett.} \textbf{62}, 2747 (1989).

\bibitem{new2}
B.-H. Chen and D.-W. Chiou,
An elementary rigorous proof of bulk-boundary correspondence in the generalized Su-Schrieffer-Heeger model,
\textit{Phys. Lett. A} {\bf 384}, 126168 (2020).
\bibitem{measure_Zak} M. Atala, M. Aidelsburger, J. T. Barreiro, D. Abanin, T. Kitagawa, E. Demler, I. Bloch, Direct measurement of the Zak phase in topological Bloch bands, \textit{Nat. Phys.} \textbf{9}, 795 (2013).

\bibitem{SupplementalMaterials} 
See the Supplemental Materials for technical details regarding the experimental setup, waveguide array design and coupling constant engineering, and additional figures.
\bibitem{Hamiltonian}
A. Peruzzo, M. Lobino, J.C.F. Matthews, N. Matsuda, A. Politi, K. Poulios, X.-Q. Zhou, Y. Lahini, N. Ismail, K. W{\"o}rhoff, Y. Bromberg, Y. Silberberg, M.G. Thompson, and J.L. O$^{\prime}$Brien, Quantum walks of correlated photons, \textit{Science} {\bf 329}, 1500 (2010).
\bibitem{review_QW}S. E. Venegas-Andraca, Quantum walks: a comprehensive review. \textit{Quantum Information Processing} \textbf{11}, 1015 (2012).

\bibitem{2DQW}  Z.-Q. Jiao, J. Gao, W.-H. Zhou, X.-W. Wang, R.-J. Ren, X.-Y. Xu,  L.-F. Qiao, X.-M. Jin, Two-Dimensional quantum walks of correlated photons. \textit{Optica} \textbf{8}, 1129-1135 (2021).
\bibitem{fs-laser_writing} A. Szameit, S. Nolte, Discrete optics in femtosecond-laser-written photonic structures. \textit{J. Phys. B} \textbf{43}, 163001 (2010).
\bibitem{2nd_coupling} F. Dreisow, A. Szameit, M. Heinrich, T. Pertsch, S. Nolte, A. T\"{u}nnermann, Second-order coupling in femtosecond-laser-written waveguide arrays, \textit{Opt. Lett.} \textbf{33}, 2689 (2008).
\bibitem{Bloch_oscilations} G. Corrielli, A. Crespi, G. Della Valle, S. Longhi, R.
Osellame, Fractional Bloch oscillations in photonics lattices, \textit{Nat. Commun.} \textbf{4}, 1555 (2013).
\bibitem{Lee} T.E. Lee, Anomalous Edge State in a Non-Hermitian Lattice,
\textit{Phys. Rev. Lett.} \textbf{116}, 133903 (2016).



	
\end{thebibliography}
\end{document}